\RequirePackage{fix-cm}
\documentclass[twocolumn,epjc3]{svjour3}  
\smartqed  % flush right qed marks, e.g. at end of proof
\RequirePackage{graphicx}
\usepackage[numbers,sort&compress]{natbib}
\usepackage{soul}
\usepackage{bm}
\graphicspath{ {plots/} }
\usepackage{latexsym} 
\usepackage{epsfig} 
\usepackage{xcolor}
\usepackage{notoccite}
\usepackage{subfigure}
\usepackage{graphicx,slashed,bbold,subfigure,amsmath,amssymb,enumerate}
\usepackage{natbib} 
\usepackage[colorlinks=true,citecolor=blue,linkcolor=blue,urlcolor=blue]{
hyperref}
\usepackage{mathtools} 
\usepackage{amsmath} 
\usepackage[normalem]{ulem}

\setcitestyle{square, numbers, comma, sort&compress} 
\newcommand{\beq}{\begin{equation}}
\newcommand{\eeq}{\end{equation}}
\newcommand{\bea}{\begin{eqnarray}}
\newcommand{\eea}{\end{eqnarray}}
\newcommand{\nn}{\nonumber}
\newcommand{\beqa}{\begin{eqnarray}}
\newcommand{\eeqa}{\end{eqnarray}}
\newcommand{\ban}{\begin{eqnarray*}}
\newcommand{\ean}{\end{eqnarray*}}
\newcommand{\bi}{\begin{itemize}}
\newcommand{\ei}{\end{itemize}}

\journalname{Eur. Phys. J. A}
\begin{document}

\title{Nambu--Jona-Lasinio $SU(3)$ model constrained by lattice QCD: thermomagnetic effects in the magnetization%\thanksref{t1}
}


\author{William R. Tavares\thanksref{e3,addr1} 
        \and
        Ricardo L. S. Farias\thanksref{e2,addr2} 
        \and
        Sidney S. Avancini\thanksref{e1,addr1}
        \and 
        Varese S. Tim\'oteo\thanksref{e4,addr3} 
        \and
        Marcus B. Pinto\thanksref{e5,addr1} 
        \and
        Gast\~ao Krein\thanksref{e6,addr4}
}

\thankstext{e1}{e-mail: william.tavares@posgrad.ufsc.br}
\thankstext{e2}{e-mail: ricardo.farias@ufsm.br}
\thankstext{e3}{e-mail: sidney.avancini@ufsc.br}
\thankstext{e4}{e-mail: varese@g.unicamp.br}
\thankstext{e5}{e-mail: marcus.benghi@ufsc.br}
\thankstext{e6}{e-mail: gastao.krein@unesp.br}


\institute{Departamento de F\'{\i}sica, Universidade Federal de Santa Catarina, 
           88040-900 Florian\'{o}polis, SC, Brazil \label{addr1}
           \and
           Departamento de F\'{\i}sica, Universidade Federal de Santa Maria, 
           97105-900 Santa Maria, RS, Brazil \label{addr2}
           \and
           Grupo de \'Optica e Modelagem Num\'erica, Faculdade de Tecnologia, 
           Universidade Estadual de Campinas, \\
           13484-332 Limeira, SP, Brazil \label{addr3}
           \and
           Instituto de F\'{\i}sica Te\'orica, Universidade Estadual Paulista, 
           Rua Dr. Bento Teobaldo Ferraz, 271 - Bloco II,  \\
           01140-070 S\~ao Paulo, SP, Brazil \label{addr4}
}

\date{Received: date / Accepted: date}

\maketitle

\begin{abstract}
We use a three-flavor Nambu--Jona-Lasinio model to study the
thermodynamics of strange quark matter under a strong magnetic
field. The model Lagrangian features flavor SU(3) four-quark
interactions and six-quark interactions that break the $U_A(1)$
symmetry. We incorporate thermomagnetic effects in the four-quark
coupling by fitting lattice results for the average of $u$ and $d$
quark condensates close to the pseudocritical temperature.
We compute the pressure at the mean
field level and obtain the magnetization of quark matter. We adopt
the recently proposed vacuum magnetic regularization (VMR) scheme,
in that divergent quark mass independent contributions are not
subtracted, thereby avoiding unphysical results for the magnetization.
We devote special attention to the renormalized magnetization,
a projected quantity that allows for direct comparisons with
lattice QCD simulations. Our results are in very good agreement
with lattice data indicating a paramagnetic behavior for
quark matter.

\end{abstract}

% % % % % % % % % % % % % % % % % % % % % % % % % % % % % % % % % % % % % % % % %
%
\section{Introduction}
\label{intro}

The possible existence of strong magnetic fields in noncentral heavy-ion
collisions~\cite{Rafelski:1975rf,Kharzeev:2007jp,Skokov:2009qp}, magnetars~\cite{Duncan:1992hi,
Kouveliotou:1998ze} and the early universe~\cite{Vachaspati:1991nm,Grasso:2000wj} 
is the topic of several recent studies. The interest is steered by the impact strong 
magnetic fields can have on prominent quantum-chromo\-dynamics (QCD) phenomena, 
notably those related to QCD's approximate chiral symmetry in the light-quark sector. 
Phenomena such as the chiral magnetic effect~\cite{Kharzeev:2007jp,Fukushima:2008xe},
chiral separation effect~\cite{Son:2004tq}, chiral Alfv\'en wave~\cite{Yamamoto:2015ria}
among several others are the subject of intense theoretical and experimental 
studies\cite{Kharzeev:2013ffa,Huang:2015oca}. The vast majority of the theoretical 
studies of such phenomena are carried out with effective models and
theories{\textemdash}Refs.~\cite{Andersen:2014xxa,Miransky:2015ava,Ayala:2021nhx} are recent 
reviews containing extensive lists of references. Such studies received a boost
when {\em ab initio} lattice QCD (LQCD) results~\cite{Bali:2011qj,Bali:2012zg} revealed 
an unexpected behavior of the chiral quark condensate as a function of the magnetic 
field strength~($B$), namely: at low temperatures the condensate increases with~$B$, 
characterizing magnetic catalysis~(MC), whereas close to the pseudocritical temperature 
of the QCD transition the condensate decreases with~$B$, characterizing inverse magnetic
catalysis~(IMC). The unexpected relates to the latter, as all effective models and earlier 
LQCD studies would predict~MC but not~IMC. The failure of earlier LQCD studies is presently
understood as being due to the use of large pion masses, much heavier than the 
physical mass~\cite{Endrodi:2019zrl,Ding:2020inp}. To incorporate the IMC effect 
within effective quark models, several ideas have been proposed~\cite{Miransky:2015ava}  
(see~\cite{Bandyopadhyay:2020zte,Andersen:2021lnk} for recent reviews). 
For instance, one simple way to conciliate LQCD predictions with those from  
effective theories, mostly in the context of those related to the Nambu--Jona-Lasinio 
model~\cite{Nambu:1961tp,Nambu:1961fr}, is to adopt a thermomagnetic dependent 
coupling~\cite{Farias:2014eca,Farias:2016gmy} constrained by the LQCD results 
of the average chiral condensates~\cite{Bali:2012zg}. Other ways to fix the 
coupling can be found in Refs. ~\cite{Ferreira:2014kpa,
Ferreira:2013tba,Endrodi:2019whh,Moreira:2020wau,Moreira:2021ety,Martinez:2018snm}. 
 
Another property observed in LQCD~\cite{Bali:2013esa,Endrodi:2013cs,Bonati:2013lca,Bonati:2013vba,Adhikari:2021bou} 
and low energy effective models~\cite{Endrodi:2013cs,Tawfik:2017cdx,Hofmann:2020lfp,Hofmann:2021bac} studies, concerns the 
paramagnetic nature of the QCD matter. Under a strong magnetic field, the response of 
QCD matter is given by the magnetization, $\mathcal{M} = \partial P / \partial (eB)$,
where $P$ is the pressure; when $\mathcal{M}>0$, one has paramagnetism. This characteristics 
can induce matter paramagnetic squeezing in non-central HICs, a phenomenon 
that can be observed in the elliptic flow $v_2$~\cite{Bali:2013owa}. We advocate that
to study physical quantities that are explicitly $B$ dependent, such as the renormalized 
magnetization~\cite{Bali:2013owa,Bali:2013esa}, apart from the incorporation of a 
thermomagnetic coupling, one also needs to adopt an adequate regularization prescription 
to avoid unphysical phenomena. A regularization method such as the well-known magnetic 
field independent regularization (MFIR) scheme~\cite{Ebert:2003yk,Ebert:1999ht,Avancini:2019wed,Duarte:2015ppa,Allen:2015paa,Menezes:2008qt,Menezes:2009uc,Avancini:2012ee,Avancini:2016fgq,Avancini:2017gck,Avancini:2018svs,Coppola:2017edn,Bandyopadhyay:2019pml},
although adequate to describe chiral transitions, leads to unphysical predictions 
for~$\mathcal{M}$. The source of the problem is that in the MFIR scheme, one also subtracts 
mass independent divergent contributions (which explicitly depend on $B$) that are 
crucial for the regularization of~$\mathcal{M}$. To avoid this problem, 
Ref.~\cite{Avancini:2020xqe} proposed a new regularization prescription, dubbed vacuum 
magnetic regularization (VMR). In that work, the VMR was proposed in the context 
of the two flavor NJL model, enforcing IMC by considering the magnetic dependent 
coupling $G(B)$ proposed in Ref.~\cite{Endrodi:2019whh}. 

In the present paper we generalize the VMR scheme to the more realistic three flavor version 
of the NJL theory. We consider a model Lagrangian that features three-flavor 
symmetry and breaks the $U_A(1)$ symmetry; the underlying theory contains four-quark
interactions with a coupling constant~$G$ and t' Hooft six-quark interactions 
with a coupling constant~$K$. In this first study, we enforce IMC by considering 
a thermomagnetic four-quark $G = G(eB,T)$ and leave the six-quark coupling~$K$ 
independent of $B$ and $T$. The present work suggests that the VMR scheme 
in conjunction with a running $G(eB,T)$ improves substantially the description
of $\mathcal{M}$ within the NJL framework. To the best of our knowledge, this work is the first study, in the
context of a flavor SU(3) NJL model, of QCD matter under a strong magnetic field 
featuring IMC and paramagnetism. We organize the presentation  
as follows. In Sec.~\ref{sec2} we present the model in the presence of a constant 
magnetic field. The same section contains the thermodynamical potential and the 
gap equations at the mean field level. Next, in Sec.~\ref{sec3}, we present the 
details of the thermomagnetic running coupling together with the renormalized 
magnetization. Numerical results are presented in Sec.~\ref{sec4}. We conclude and
discuss perspectives of future work in Sec.~\ref{sec5}. 
 
% % % % % % % % % % % % % % % % % % % % % % % % % % % % % % % % % % % % % % % % %
%
\section{The model}
\label{sec2}

There exist several options for NJL-type of Lagrangians that feature three-flavor 
symmetry and break the unwanted $U_A(1)$ symmetry~\cite{Vogl:1991qt,
Klevansky:1992qe,Hatsuda:1994pi}. We use the one first written down in 
Ref.~\cite{Kunihiro:1987bb}; it comprises flavor $U_L(3)\otimes U_R(3)$ 
symmetric four-fermion interactions and a~six-point interaction that 
breaks the $U_A(1)$ symmetry, given by the 't Hooft determinant:
\begin{eqnarray}
\mathcal{L}_{sym} &=& G \sum_{a=0}^8 \left[  \left( \bar{\psi} \lambda_a \psi  \right)^2 +
\left( \bar{\psi} i\gamma _{5 }\lambda_a \psi  \right)^2 \right], 
\label{Lsym} \\
\mathcal{L}_{det} &=&   -K \left[ \det \bar{\psi} ( 1+ \gamma_5) \psi  +
\det \bar{\psi} ( 1- \gamma_5) \psi \right].
\label{Ldet}
\end{eqnarray}
Here, $\psi$ represents the three-flavor multiplet of Dirac spinors 
$\psi = (\psi_u,\psi_d,\psi_s)^T$, and $\lambda^a, a= 1, \cdots, 8$ are the 
$SU(3)$ Gell-Mann matrices and $\lambda^0 = \sqrt{3/8}\,I$, with $I$ the $3\times 3$ 
unit matrix. The determinant is in flavor space, it can be written in terms of the 
Levi-Civita tensor $\epsilon_{ijk}$ as $\det \bar\psi {\cal O}\psi = \sum_{i,j,k} 
\epsilon_{ijk} (\bar\psi_u {\cal O} \psi_i) (\bar\psi_d {\cal O} \psi_j) 
(\bar\psi_s {\cal O} \psi_k)$ with $i,j,k=u,d,s$ and  ${\cal O} = (1\pm\gamma_5)$.     
The complete Lagrangian density includes an explicit symmetry-breaking mass term and 
the coupling of an electromagnetic field \cite{Menezes:2009uc}: 
\begin{eqnarray}
\mathcal{L}=\bar{\psi} \left[ i\gamma _{\mu } D^{\mu }-\hat{m} \right] \psi 
+ \mathcal{L}_{sym} +\mathcal{L}_{det} - \frac{1}{4}
{F}^{\mu \nu}{F}_{\mu \nu},
\label{lag}
\end{eqnarray}
where $D^{\mu } = \partial^{\mu }~+~i Q A^{\mu }$, with $A^\mu$ being the 
electromagnetic gauge field potential and ${F}_{\mu \nu}$ the corresponding 
field tensor, $Q$ is the quark charge matrix $Q = {\rm diag}(q_u, q_d, q_s) 
= {\rm diag}(2/3\, e,-1/3\,e,-1/3\,e)$ where $e=\sqrt{4\pi/137}$ is the elementary electric charge, 
and $m$ the current-quark mass matrix $m = {\rm diag}(m_u,m_d,m_s)$. We choose a 
spatially uniform, time independent magnetic field of strength~$B$ pointing in 
the $\hat{\boldmath{z}}$ direction, so that one can can choose the gauge field 
as $A^{\mu} =  x  B \delta^{\mu 2}$. 

The mean-field grand-canonical potential $\Omega\left(T,eB \right)$ of the model was computed 
in Ref.~\cite{Menezes:2009uc} within the MFIR prescription, but one can readily transcribe 
that derivation to the VMR prescription~\cite{Avancini:2020xqe}. In both prescriptions,
the generic form of $\Omega\left(T,eB \right)$ can be written as 
\begin{equation}
\Omega(T,eB) = \sum_f \omega_f + 2G \sum_f \phi_f^2 - 4K ~  \phi_u\phi_d\phi_s, 
\label{omega}
\end{equation}
where $\omega_f,\; f=u,d,s$ comes from the first term in the Lagrangian and 
$\phi_f$ is the flavor $f$ quark condensate that comes from the quark-quark interactions. 
Both $\omega_f$ and $\phi_f$ can written as sum of terms with well defined origin:
vacuum, purely magnetic and medium (mixed temperature and magnetic field). The differences
between the MFIR and VMR appear in the expression for $\omega_f$.
Specifically, the VMR $\omega_f$ can be written~as~\cite{Avancini:2020xqe}:
\begin{eqnarray}
\omega_f = \omega^{vac}_f~ +~ \omega^{mag}_f~+ ~\omega^{field}_f~ +~ \omega^{med}_f,
\label{omega_f}
\end{eqnarray}
where
\begin{eqnarray}
\omega_f^{vac}~&=&~  \frac{N_c}{8\pi^2} \left[ M_f^4 \log\left(\frac{\Lambda 
+ \epsilon_f}{M_f}\right) - \epsilon_f \Lambda (\Lambda^2+\epsilon_f^2) \right], 
\label{om_vac} \\
\omega_f^{mag}~&=&~ - \frac{N_c(|q_f|B)^2}{2\pi^2} \left[\zeta^\prime (-1,x_f) 
+  \frac{1}{4}x_f^2  \right. \nn \\
&& \left. - \, \frac{1}{2}(x_f^2-x_f)\log x_f 
- \frac{1}{12}\left(1 + \log x_f\right) \right],
\label{om_mag} \\
\omega_f^{field}~&=&~ -\frac{N_c(|q_f|B)^2}{24\pi^2}\log\left(\frac{M_f^2}{\Lambda^2}\right),
\label{om_field} \\
\omega_f^{med}~&=&~  - T \sum_{k=0}^{\infty} \left(2-\delta_{0k}\right) 
\frac{|q_f| B}{2\pi^2} \nn \\
&& \times \, \int_{-\infty}^{\infty} dp ~ \ln \left( 1 +  e^{-E_f/T} \right),
\label{ommed2} 
\end{eqnarray}

\noindent where $\zeta^\prime(-1,x_f)=\frac{d\zeta(z,x_f)}{dz} |_{z=-1}$.

The MFIR expressions differ from the above by the absence of the last term in
Eq.~(\ref{om_mag}) and the $\omega_f^{field}$ contribution in Eq.~(\ref{om_field}).
As mentioned, those terms are essential for describing the SU(2) lattice data.
In the above expressions, we defined $\epsilon_f = (\Lambda^2 + M_f^2)^{1/2}$ 
where $\Lambda$ is a three-dimensional cutoff,  
$x_f = {M_f^2}/{2|q_f|B}$, $\zeta(s,x)$ is the Riemann-Hurwitz zeta function, 
and $E_f= (p^2+M_f^2+2|q_f|B k)^{1/2}$. In the previous definitions, $M_f$ represents the flavor~$f$ constituent quark mass, 
determined by the
familiar NJL gap equations: 
\begin{eqnarray}
 M_u &=& m_u-4G\phi_u+2K\phi_d \phi_s, \label{gap-u}\\
 M_d &=& m_d-4G\phi_d+2K\phi_u \phi_s, \label{gap-d}\\
 M_s &=& m_s-4G\phi_s+2K\phi_u \phi_d . \label{gap-s}
\end{eqnarray}
In this first study only the coupling $G$ is enforced to be $T$ and $B$ dependent.  
As mentioned above, the flavor~$f$ quark condensate can also be expressed as
a sum of vacuum, magnetic field and medium contributions, namely:
\begin{eqnarray}
\phi_f=\phi_f^{vac}+\phi_f^{mag}+\phi_f^{med},
\label{cond1}
\end{eqnarray}
where 
\begin{eqnarray}
\phi_f^{vac}~&=&~  - \frac{N_c M_f}{2\pi^2} \left[ \Lambda \epsilon_\Lambda 
-M_f^2  \log \left( \frac{\Lambda + \epsilon_\Lambda}{M_f} \right) \right],
\label{phi-vac} \\
\phi_f^{mag}~&=&~  - \frac{M_f N_c |q_f|B}{2\pi^2} \left[ \log \Gamma (x_f) 
- \frac{1}{2}\log(2\pi)
+ x_f \right. \nn \\
&& \left. - \, \frac{1}{2} (2x_f-1)\log x_f \right],
\label{phi-mag} \\
\phi_f^{med}~&=&~ \sum_{k=0}^{\infty} \left(2-\delta_{0k}\right)  
\frac{|q_f| B~ N_c M_f}{2\pi^2} \nn \\
&& \times \, \int_{-\infty}^{\infty} dp \, \frac{1}{E_f} \, \frac{1}{e^{E_f/T}+1}, 
\label{phi-med}
\end{eqnarray}
where $\Gamma$ is the Euler gamma function. 

% % % % % % % % % % % % % % % % % % % % % % % % % % % % % % % % % % % % % % % % %
%
\section{Running coupling and magnetization}
\label{sec3}

We use the average quark condensate $(\Sigma_u+\Sigma_d)/2$~\cite{Bali:2012zg}  
to adjust the running coupling $G(eB,T)$. To this end, we adopt the following
parametrization of the Gell-Mann-Oakes-Renner (GOR) 
relation~\cite{Bali:2012zg,Farias:2016gmy} 
\begin{eqnarray}
\Sigma_f = \frac{2m}{m_{\pi}^2f_{\pi}^2}
\left[ \left \langle \bar{\psi}_f\psi_f \right \rangle_{B,T} 
- \left \langle \bar{\psi}_f\psi_f \right \rangle_{00} \right] + 1, 
\label{GOR}
\end{eqnarray}
where $\langle \bar{\psi}_f\psi_f \rangle_{00}^{1/3}=-230.55$ MeV is the quark 
chiral condensate at $B=T=0$, $f_{\pi} =86$ MeV, $m_{\pi} = 135$~MeV, and $m=5.5$ MeV. Adopting this particular set of phenomenological values will allow us to perform direct comparisons with  the  lattice results of Ref. \cite {Bali:2012zg}.

The $T$ and $B$ dependent condensates $\langle \bar{\psi}_f\psi_f \rangle_{B,T} = \phi_f$ 
are evaluated within the NJL expressions given in the previous section, 
Eqs.~(\ref{cond1})-(\ref{phi-med}). The constituent quark masses $M_u, M_d$, and 
$M_s$ in the expressions for $\phi_f$ are obtained by solving the gap
equations Eqs.~(\ref{gap-u})-(\ref{gap-s}). The $T$ and $B$ running of the coupling $G=G(eB,T)$
is dictated by an {\it ansatz} similar to that used for the SU(2) model~
\cite{Farias:2016gmy}, namely:
\begin{equation}
G(eB,T) = \alpha(eB)\left(1-\frac{d(eB)}{1+ e^{\beta(eB)(T_a(eB)-T)}}\right)+s(eB). 
\label{GB}
\end{equation}

\noindent This expression has been adopted for mere convenience since it is well adapted for the adjustment of LQCD results. Of course, other possibilities may be used with similar results.

\begin{table}[htb]
\caption{Values of parameters in Eq.(\ref{GB}) in appropriate GeV units;
$d(B)$ is dimensionless.}
 \begin{tabular}{cccccc} 
 \hline\noalign{\smallskip}
 $eB$ & $\alpha(eB)$ & $\beta(eB)$ & $T_a(eB)$ & $d(eB)$ & $s(eB)$ \\ 
 \noalign{\smallskip}\hline\noalign{\smallskip}
 0.0 & 2.1534 & 420.95 & 0.1678   & 0.3506 & 2.0793 \\[0.05cm] 
 0.2 & 1.7571 & 142.44 & 0.1844   & 1.4636 & 2.3358 \\[0.05cm]
 0.4 & 0.8158 & 183.46 & 0.1712   & 2.0641 & 2.8016 \\[0.05cm]
 0.6 & 0.7148 & 128.16 & 0.1720 & 3.2874 & 2.3080 \\
\noalign{\smallskip}\hline
\end{tabular}
\label{tab}
\end{table}

Table \ref{tab} displays the numerical values for the parameters appearing in Eq.~(\ref {GB}). 
These selected values are those which best fit the average quark condensate. 
Since Ref. \cite{Bali:2012zg} offers no data points between $T\in [0, 113 \text{MeV}]$,  
we follow the strategy of Ref.~\cite{Farias:2016gmy}, in that we fit~$G$ using the 
available lattice data to extrapolate the results to lower temperatures. 

In the next section we present the fitting results of the lattice data 
for $(\Sigma_u+\Sigma_d)/2$~\cite{Bali:2012zg} and compare the predictions 
of the model for the pseudocritical temperature with the corresponding 
lattice results. We will also show results for the renormalized 
magnetization, $\mathcal{M}^r$~\cite{Avancini:2020xqe,Bali:2013esa}:

\begin{eqnarray}
\mathcal{M}^r \cdot eB = \mathcal{M} \cdot eB - (eB)^2 \lim_{eB\rightarrow 0} 
\frac{\mathcal{M} \cdot eB}{(eB)^2}\bigg| _{T=0},\label{Mr}
\end{eqnarray}

\noindent where $\mathcal{M} = - {\partial \Omega(T,eB)}/{\partial (eB)}$, with $\Omega(T,eB)$ 
representing the thermodynamical potential, Eq.(\ref{omega}). Although the NJL 
model is a nonrenormalizable field theory, this prescription, which motivated the 
VMR for the SU(2) NJL model,  gives us the possibility to compare our results 
directly with the LQCD data{\textemdash}see Ref.~\cite{Avancini:2020xqe} for more 
details. 

% % % % % % % % % % % % % % % % % % % % % % % % % % % % % % % % % % % % % % % % %
%
\section{Numerical results}\label{sec4}

The following set of parameters is adopted in this work: $\Lambda = 631.4$ MeV, 
$m_u=m_d=5.5$ MeV, $m_s=135.7$ MeV, $G=1.835/\Lambda^2$ and 
$K=9.29/\Lambda^5$~\cite{Hatsuda:1994pi}. We note that the value of $G$ corresponds 
to the vacuum value, not the extrapolated $G(B=0,T=0)$.

\begin{figure}[h]
\begin{center}
\vspace{0.5 cm}
\includegraphics[width=0.45\textwidth]{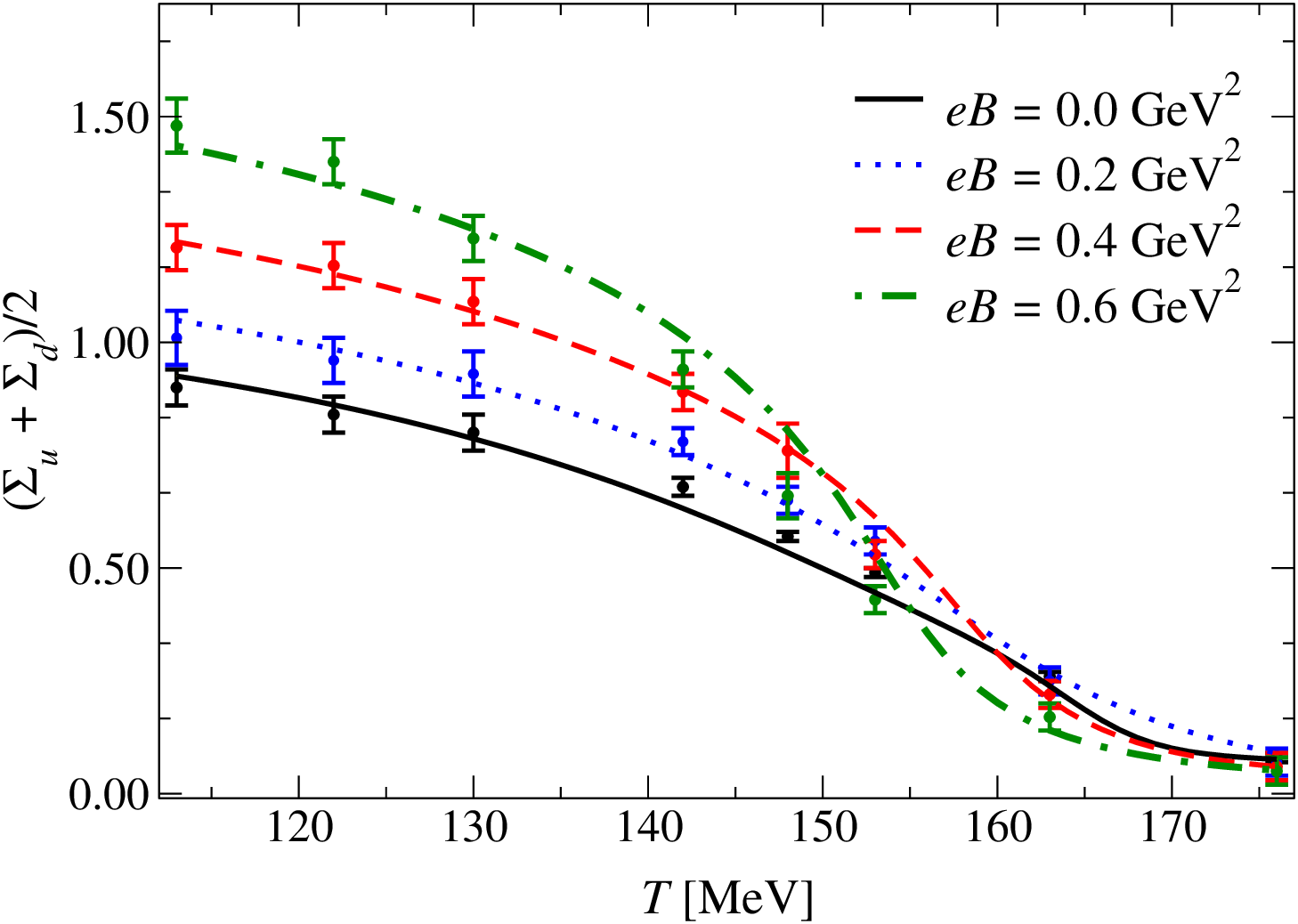}
\end{center}
\caption{Fit of the LQCD data~\cite{Bali:2012zg} for the average quark condensate as a function 
of the temperature for several magnetic field values.}
\label{cond}
\end{figure}

\begin{figure}[h]
\begin{center}
\includegraphics[width=0.45\textwidth]{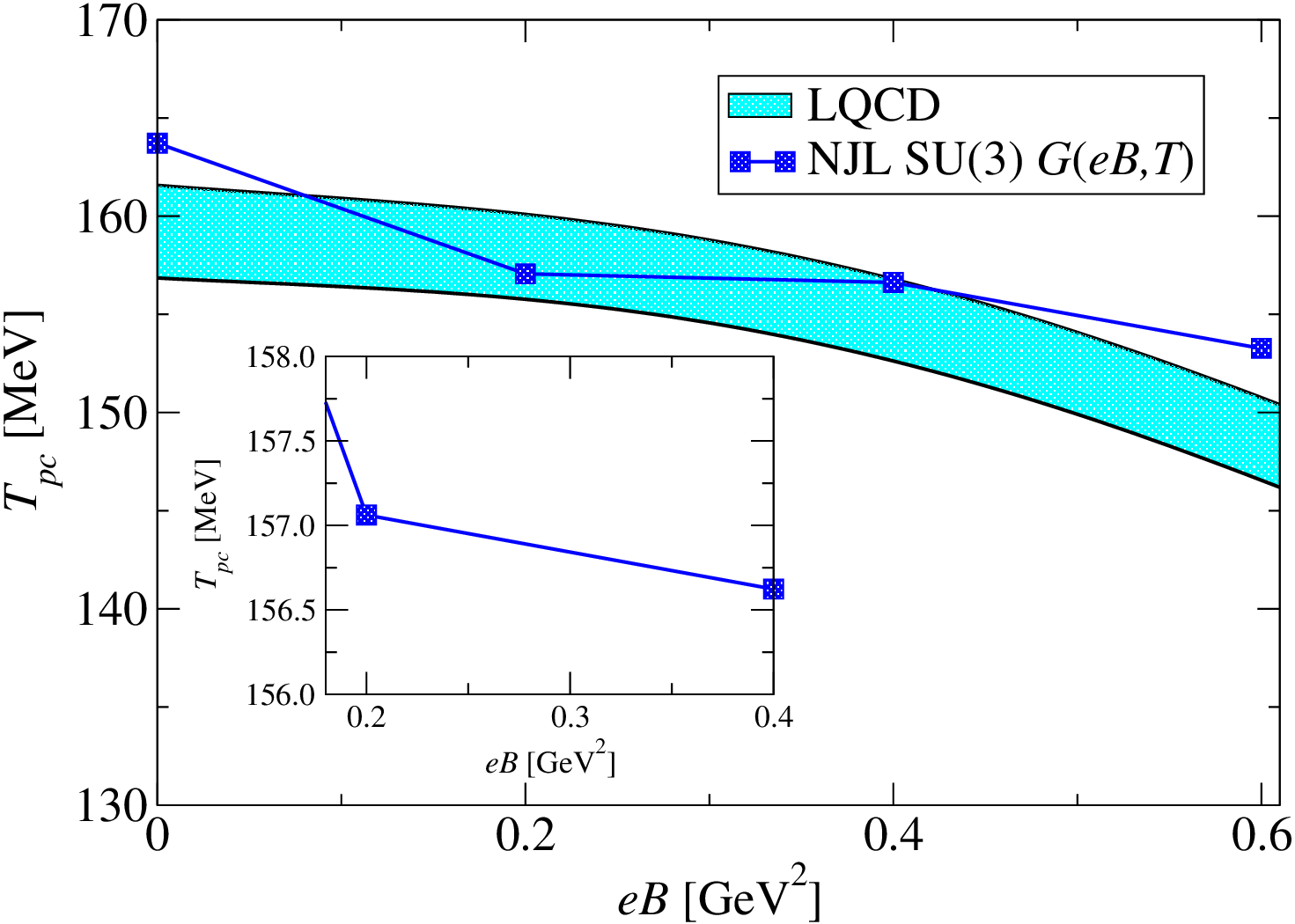}
\end{center} 
\caption{Pseudocritical temperature as a function of the magnetic field compared 
with LQCD results in the extrapolation limit (blue band)~\cite{Bali:2011qj}.}
\label{TcvsB}
\end{figure}

In figure~\ref{cond} we show the thermal dependence of the average quark condensate 
for different values of $B$ using the thermomagnetic dependent coupling $G(eB,T)$.
The fit to the lattice data is very good. The figure clearly displays the IMC 
phenomenon for $eB\gtrsim 0.2 \text{GeV}^2$ and $T \gtrsim 150 \text{MeV} $. These 
results can be better understood with the aid of figure~\ref{TcvsB}, which displays the 
predicted pseudocritical temperature $T_{pc}$ as a function of~$eB$.
The inset in this figure shows that $T_{pc}$
decreases with $eB$ within the range $0.2\, \text{GeV}^2 < eB < 0.4 \,\text{GeV}^2$. The predictions 
of the model compare fairly well with the lattice data in the continuum extrapolation 
limit (blue band)~\cite{Bali:2011qj} within the range of magnetic fields considered.

\begin{figure}
\begin{center}
\includegraphics[width=0.45\textwidth]{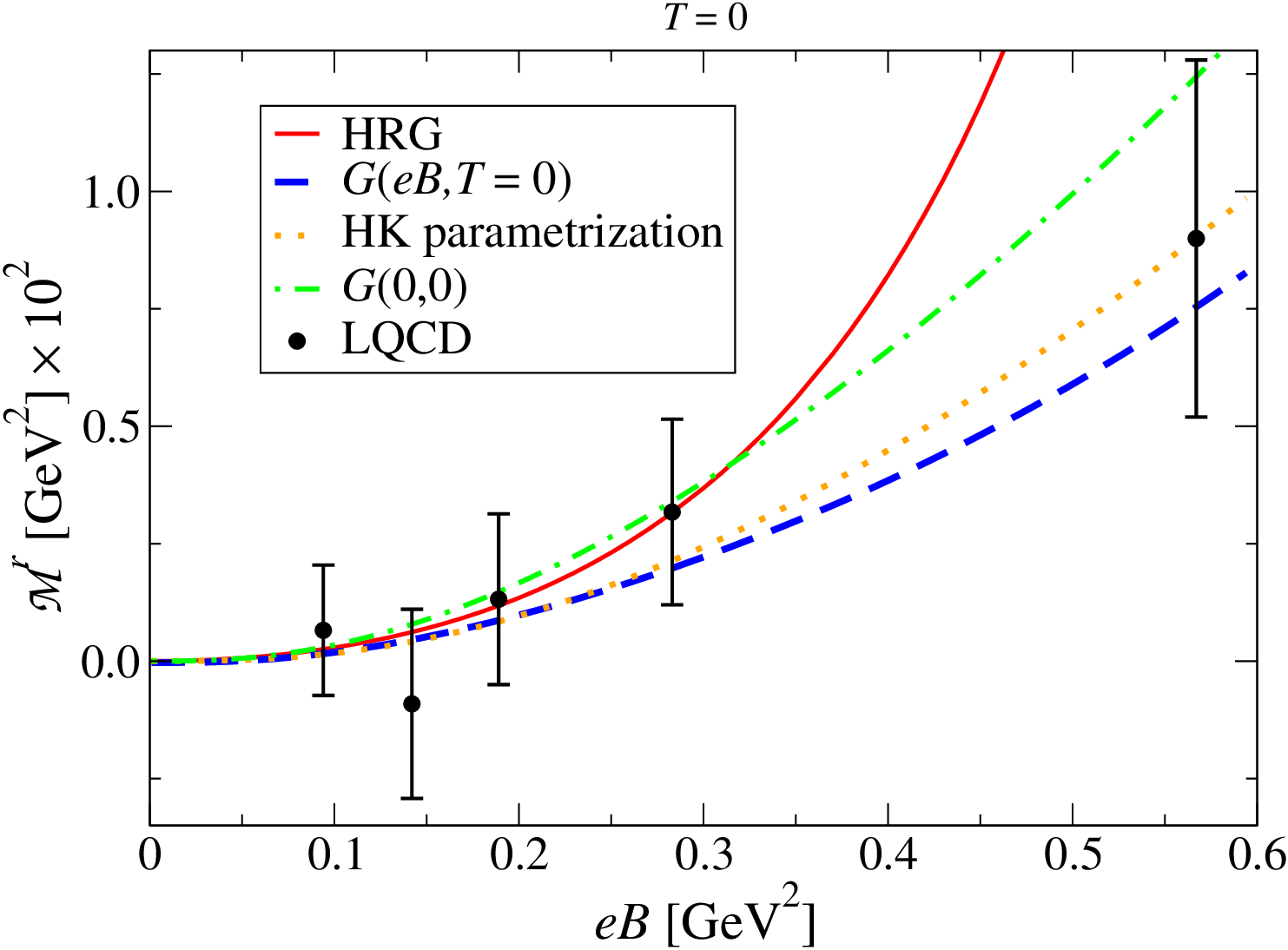}\vspace{1.0 cm}
\includegraphics[width=0.45\textwidth]{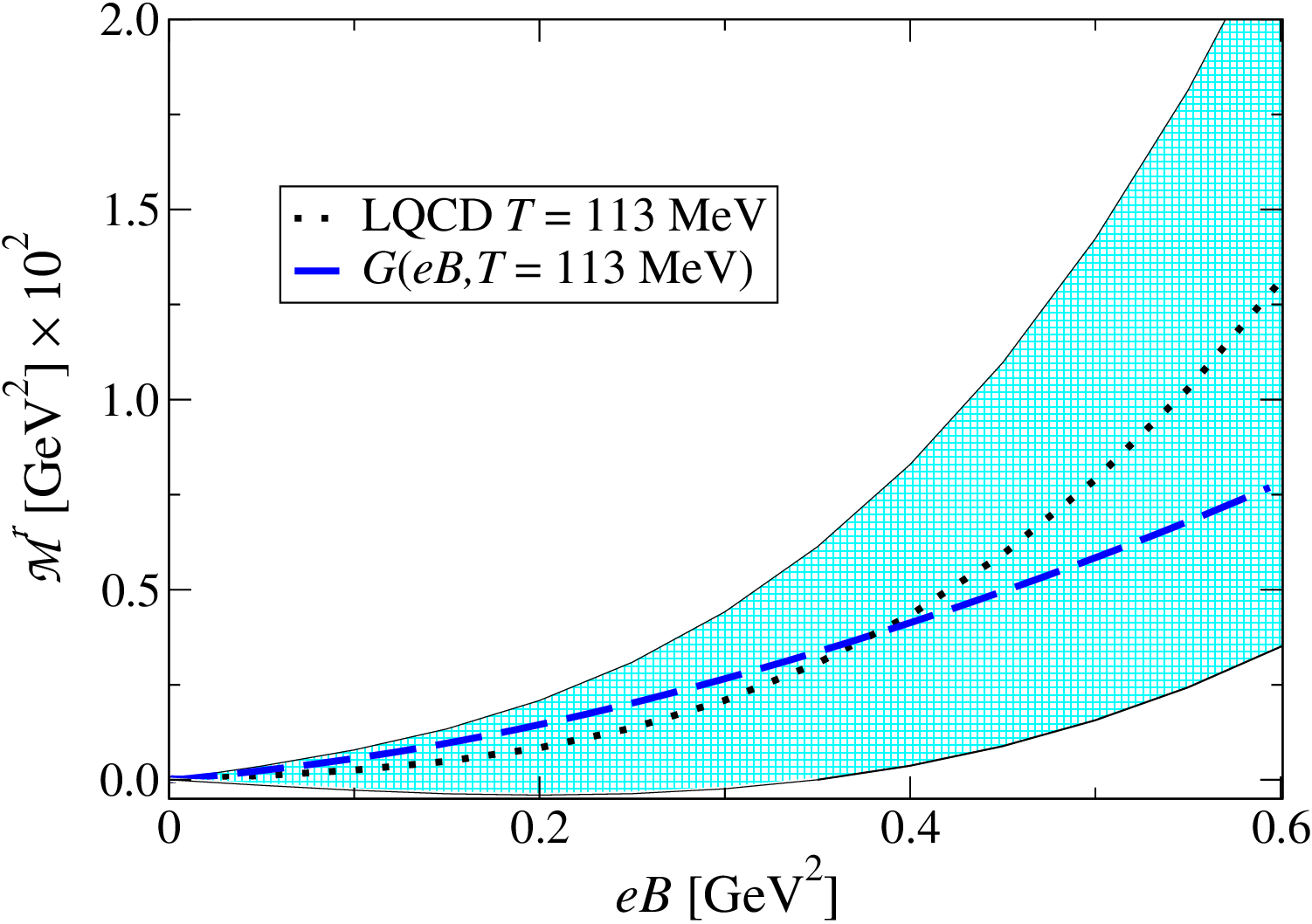}\vspace{1.0 cm}
\includegraphics[width=0.45\textwidth]{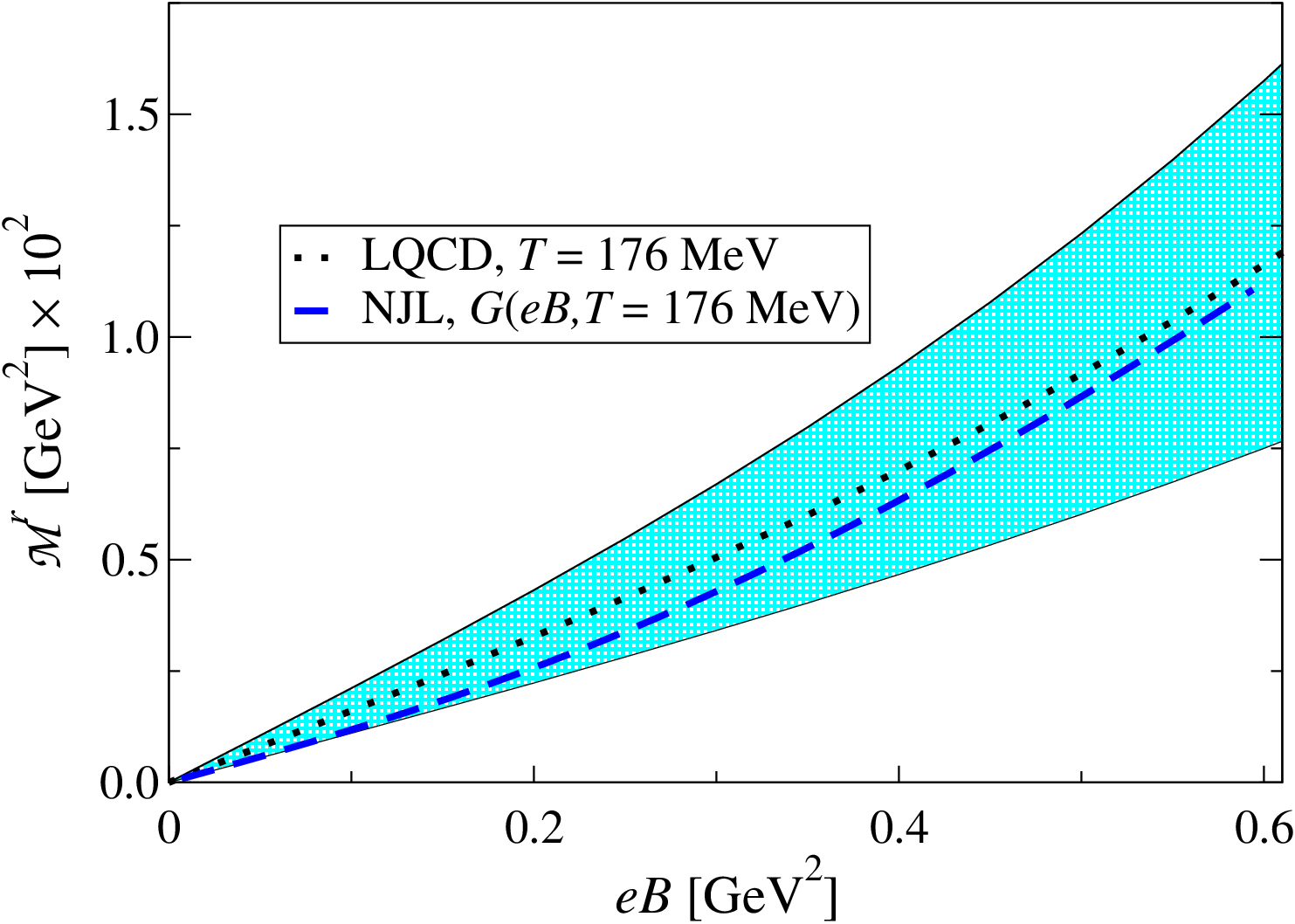}
\end{center}
\caption{Renormalized magnetization as function of magnetic field for
three sets of temperatures: $T=0$ (top panel), $T=113$~MeV (center panel) and $T=176$~MeV (bottom panel).
The blue bands represent the error bands of the LQCD results black 
dotted line represents the fit for the LQCD results Ref.~\cite{Bali:2013owa}).}
\label{fig:Mr}
\end{figure}

Figure~\ref{fig:Mr} displays the magnetic field dependence of the renormalized 
magnetization $\mathcal{M}^r$ for three sets of temperature values. The top panel, 
for $T=0$, shows results for three of the coupling values: $G(eB,T=0)$ (blue dashed),
$G=1.835/\Lambda^2$ (yellow dotted)~\cite{Hatsuda:1994pi}, and $G(0,0)$ 
(green dot-dashed){\textemdash}the six-quark coupling is the same in
all cases, $K=9.29/\Lambda^5$~\cite{Hatsuda:1994pi}. The first two $G$-coupling
values lead to fairly good agreement with the LQCD data of Ref.~\cite{Bali:2013esa}, 
whereas the agreement with the coupling $G(0,0)$ is only good up to $eB \simeq 0.3~{\rm GeV}^2$. 
The figure also displays predictions from the hadron resonance gas (HRG) 
model~\cite{Bali:2013owa}, which also agree with LQCD data up to $eB \simeq 0.3~{\rm GeV}^2$ only. 
In the other two panels, we show results obtained with $G(eB,T)$ for 
$T~=~113$~MeV (center panel) and $T~=~176$~MeV (bottom panel). Very good agreement 
with the LQCD data of Ref.~\cite{Bali:2013owa} is again observed for
these temperatures. 

\begin{figure}[h]
\begin{center}
\includegraphics[width=0.45\textwidth]{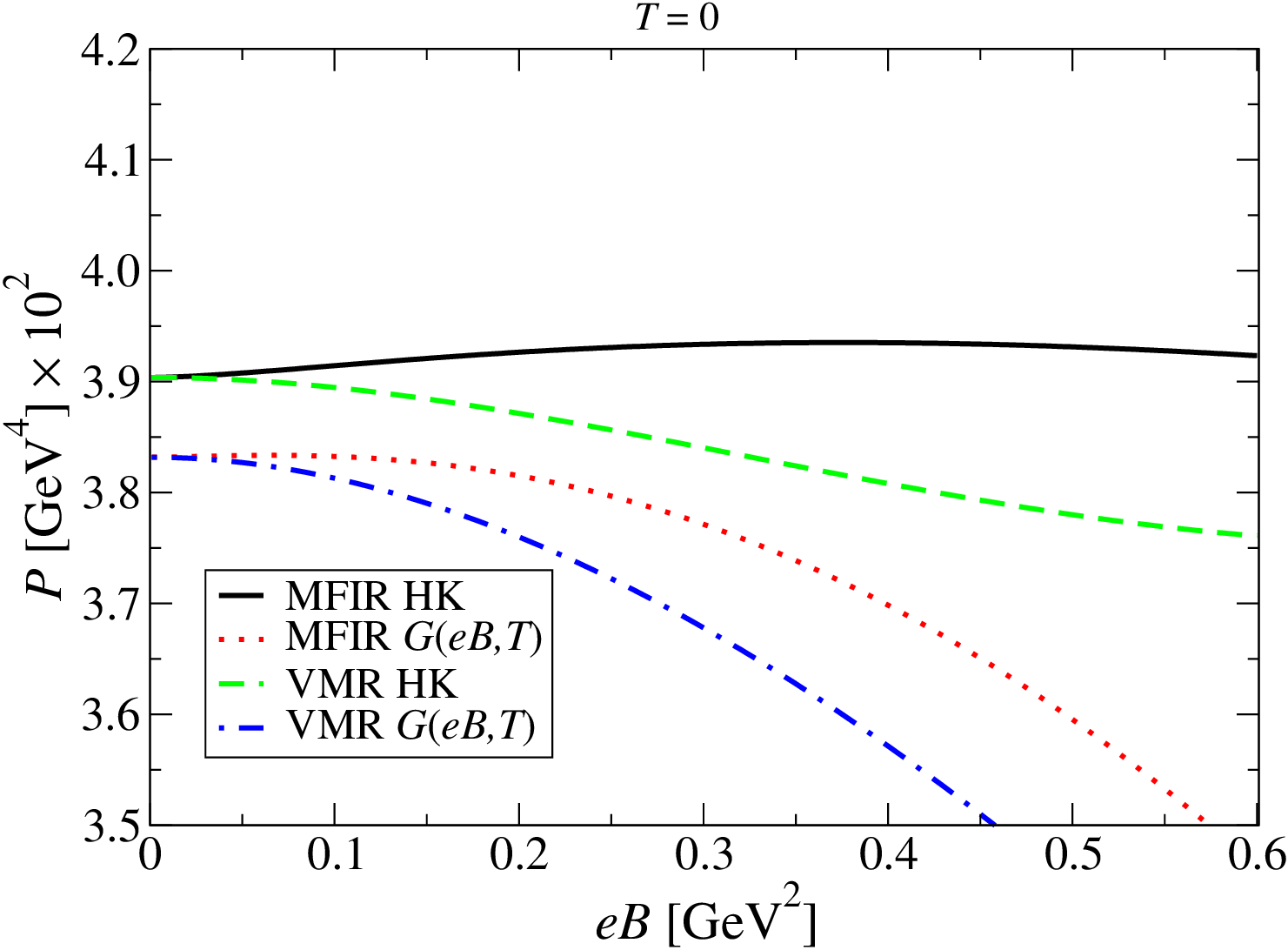}\vspace{1.0 cm}
\includegraphics[width=0.45\textwidth]{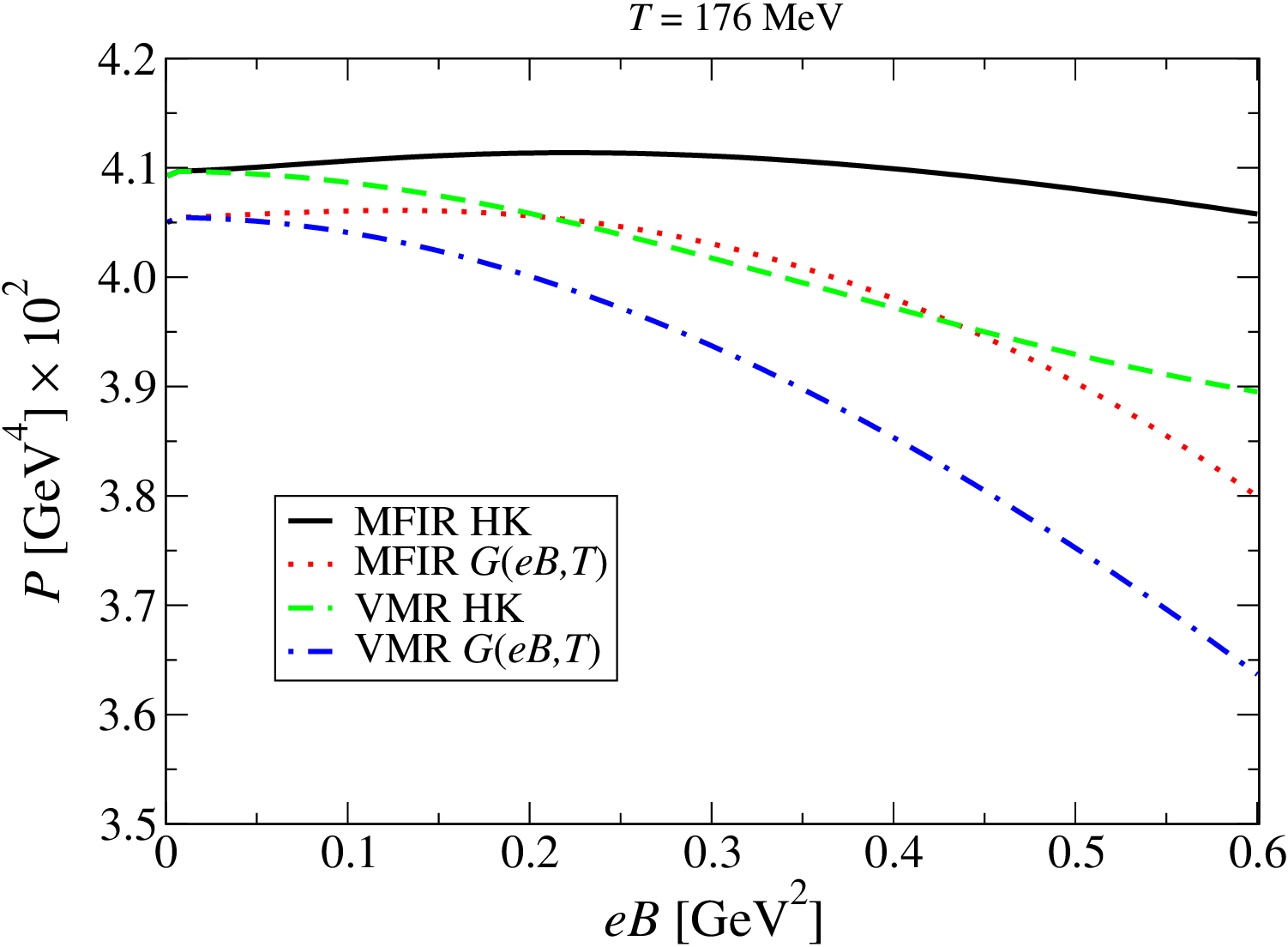}
\end{center}
\caption{Pressure ($P = - \Omega(T,eB)$) as function of the magnetic field for 
the temperatures $T=0$ (top panel) and $T=176$ MeV (bottom panel) in the MFIR and VMR schemes. 
Results obtained with $G(eB,T)$ and the vacuum coupling~$G$ (denoted HK)~\cite{Hatsuda:1994pi} 
(in both cases, the coupling $K$ is the same). 
}
\label{pressure}
\end{figure}

For completeness, we present in figure~\ref{pressure} the model's predictions 
for the pressure, $P = - \Omega(T,eB)$, obtained in the MFIR and VMR schemes. 
The figure displays the magnetic field dependence of~$P$ for $T=0$ (top panel) 
and $T~=~ 176$~MeV (bottom panel). We do not show the results for $T=113$ MeV, 
the temperature explored in figure~\ref{fig:Mr}, because they are very similar 
to those at~$T=0$. We compare results obtained with the running coupling, 
$G(eB,T)$, as well as with the fixed value of Ref.~\cite{Hatsuda:1994pi}, indicated 
by HK in the figure{\textemdash}again, the six-quark coupling~$K$ is the same used above. 
The figure reveals that the pressure values predicted using $G(eB,T)$ are systematically 
lower than those where a $T$ and $B$ independent~$G$ has been used. This behavior is observed 
for both temperature values considered. Interestingly, the MFIR and VMR predictions 
have qualitatively different $eB$ dependence for both values of~$T$, a feature
already pointed out in Ref.~\cite{Avancini:2020xqe} for the SU(2) case: in the MIFR 
scheme, the $eB$ dependence is nonmonotonic, $P$~starts increasing and then decreases, 
whereas in the VMR scheme, $P$ decreases monotonically with~$eB$. This feature is 
observed for both temperature and coupling sets used. These results evince, now 
also for the SU(3) case, how the (divergent) mass independent terms present in the 
VMR scheme affect the $B$ dependence of the pressure. 

\begin{figure}[h]
\begin{center}
\vspace{0.5 cm}
\includegraphics[width=0.45\textwidth]{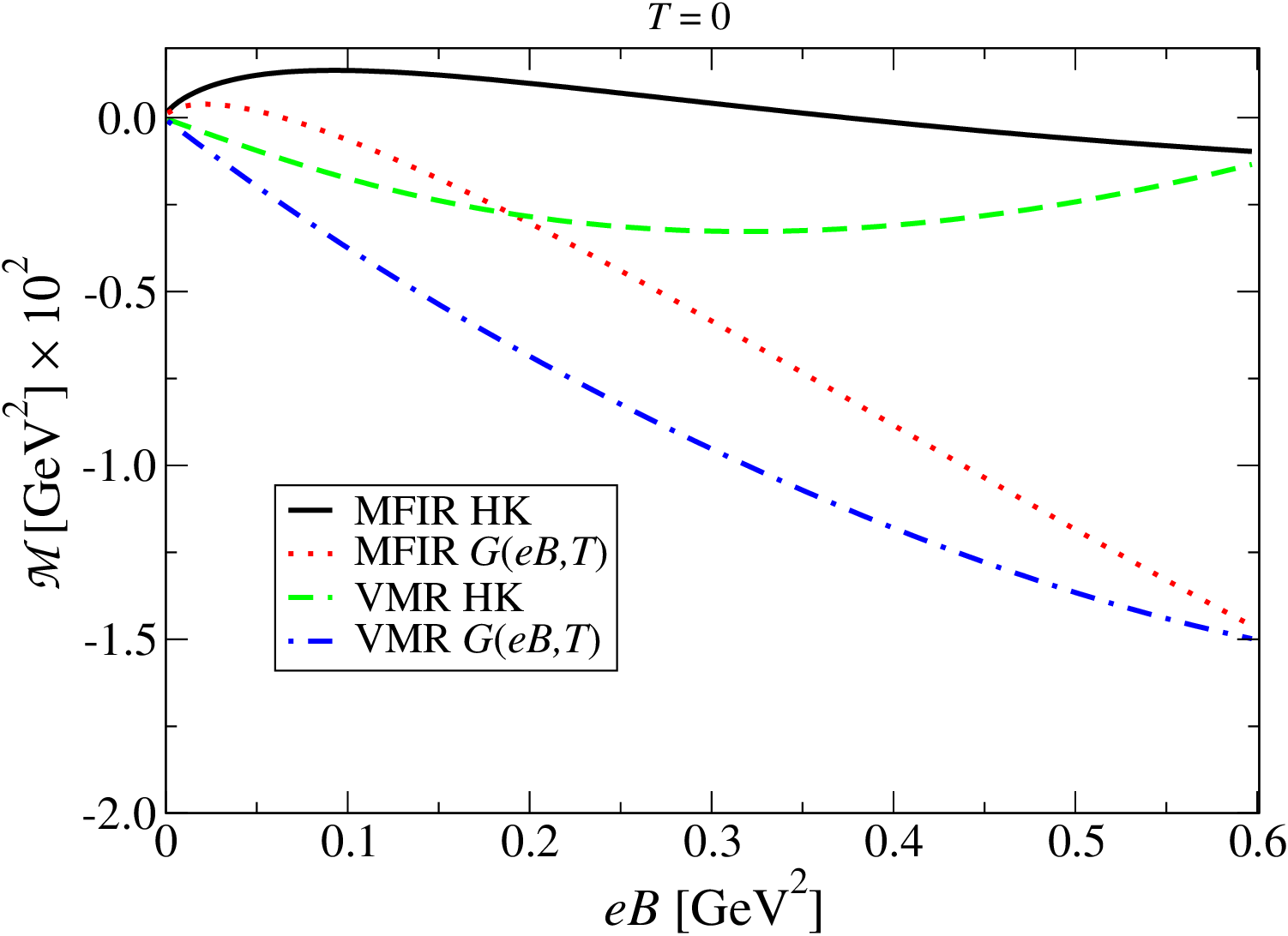}\vspace{1.0 cm}
\includegraphics[width=0.45\textwidth]{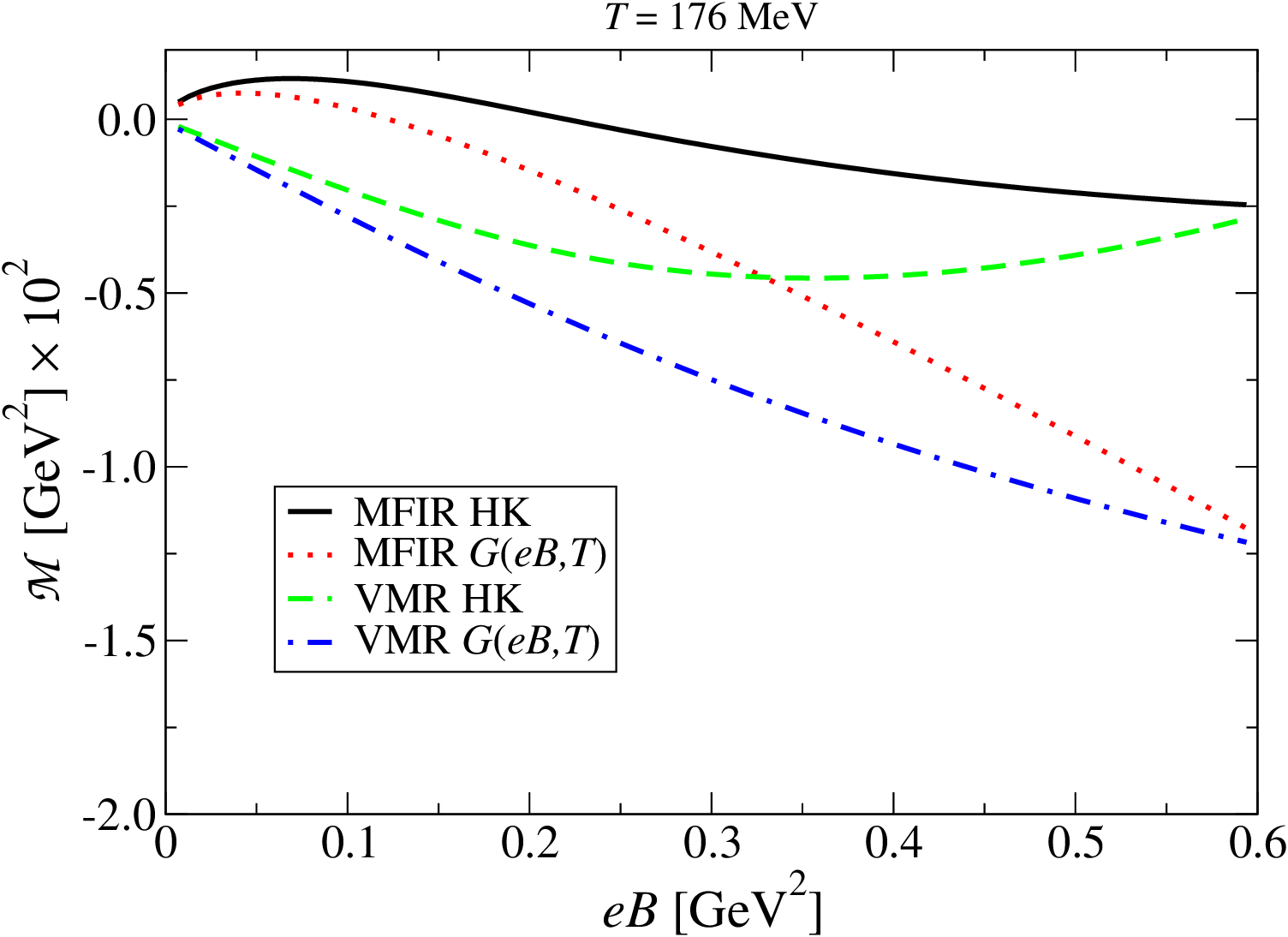}
\end{center}
\caption{Magnetization as function of the magnetic field for the fixed temperatures $T=0$ 
(top panel) and $T=176$ MeV (bottom panel).}
\label{Mag}
\end{figure}

The importance of the mass independent terms pre\-sent within the VMR can be further  highlighted by  examining the magnetization displayed in Fig. \ref{Mag}.
The figure shows the results at $T=0$ case (top panel)  and  $T=176$ MeV (bottom panel) for both schemes with a fixed and a running four fermion coupling.
One can easily see that in the MFIR scheme we have $\mathcal{M}>0$ for $eB \lesssim 0.3 \text{GeV}^2$ in $T=0$ and $eB \lesssim 0.2 \text{GeV}^2$ in $T=176$ MeV with fixed coupling. 
The MFIR results with $G(eB,T)$ show a similar behavior  but $\mathcal{M}>0$  is only observed at rather  low $B$ values. The VMR scheme shows $\mathcal{M}<0$ for a fixed coupling at both temperatures considered 
(although the magnetization at these two temperature values increases when $T\gtrsim 0.3 \text{GeV}^2$). The VMR scheme, with $G(eB,T)$, predicts a more 
dramatic (and completely monotonic) decrease of $\mathcal{M}$ as the magnetic field increases. Highlighting, once again, the crucial role played by contributions which are subtracted within the MFIR method.

% % % % % % % % % % % % % % % % % % % % % % % % % % % % % % % % % % % % % % % % %
%
\section{Conclusions}\label{sec5}

In this work we extended the recently  \cite{Avancini:2020xqe} proposed VMR
scheme to describe magnetized strange quark matter within a
three flavor NJL model framework. The thermomagnetic running of the four fermion coupling, $G(eB,T)$, was determined by fitting lattice QCD data for the
quark condensate, reproducing in this way the inverse magnetic catalysis effect predicted by most lattice evaluations. When regulated with the VMR the thermodynamical potential presents  mass independent terms which are usually subtracted in other schemes such as the  MFIR. Since these extra terms are strongly $B$ dependent their presence greatly impacts physical observables such as  the magnetization. Despite the fact that the four dimensional NJL represents a nonrenormalizable theory one may, nevertheless, define a projected quantity, $\mathcal{M}^r$, which allows for a direct comparison with   LQCD results in a satisfactory way (see Ref. \cite{Avancini:2020xqe} for details). Our results indicate a very good concordance with LQCD results with and
without the thermomagnetic four fermion scalar coupling although IMC was present just when $G(eB,T)$ is considered, as expected. To show the importance of 
the mass independent terms in the thermodynamical potential, we have also compared the pressure and the magnetization as
$\mathcal{M}=\partial P / \partial (eB)$ as function of the magnetic field  in both MFIR  and VMR schemes. The comparison with LQCD data has shown that the  MFIR procedure is more sensitive to the variations of the magnetic field furnishing less reliable results.
The present work shows that the VMR, originally proposed in the context of two flavors,  can be readily generalized to the more realistic three flavor case. We have also demonstrated that when this regularization method is used in conjunction with an adequate thermomagnetic four fermion coupling the effective model is able to produce results which are in line with LQCD predictions. These include the IMC phenomenon as well as the paramagnetic character of the quark matter. A more complete analysis with the inclusion of the thermomagnetic
dependence of the six fermion t'Hooft coupling is underway.

\begin{acknowledgements}
This work was partially supported by Conselho Nacional de Desenvolvimento Cient\'ifico 
e Tecno\-l\'o\-gico  (CNPq), Grants No. 309598/2020-6 (R.L.S.F.), No. 304518/2019-0 (S.S.A.) 
No. 303846/2017-8 (M.B.P), and No. 309262/2019-4 (G.K.), No. 306615/2018-5 (V.S.T.); Coordena\c c\~{a}o  de 
Aperfei\c coamento de Pessoal de  N\'{\i}vel Superior - (CAPES) Finance  Code  001 ( W.R.T); 
Funda\c{c}\~ao de Amparo \`a Pesquisa do Estado do Rio 
Grande do Sul (FAPERGS), Grants Nos. 19/2551- 0000690-0 and 19/2551-0001948-3 \\(R.L.S.F.);
Funda\c{c}\~ao de Amparo \`a Pesquisa do Estado de S\~ao Paulo (FAPESP), Grant No. 
2018/25225-9 (G.K.), No. 2019/10889-1 (V.S.T.); Fundo de Apoio ao Ensino, Pesquisa e \`a Extens\~{a}o (FAEPEX), Grant No. 3258/19 (V.S.T.). 
The work is also part of the project Instituto Nacional de Ci\^encia 
e Tecnologia - F\'isica Nuclear e Aplica\c{c}\~oes (INCT - FNA), Grant No. 464898/2014-5. 

\end{acknowledgements}

\bibliographystyle{spphys}       % APS-like style for physics
\bibliography{paperEPJA_njl_su3_bib.bib}   % name your BibTeX data base

%\begin{thebibliography}{}
% etc
%\end{thebibliography}

\end{document}